\documentclass[twocolumn,showpacs,preprintnumbers,amsmath,amssymb]{revtex4}
\usepackage{epsfig}
\usepackage{amsmath}
\usepackage{times}

\begin{document}

\title{Condensation in a zero range process on weighted scale-free networks}
\author{Ming Tang, Zonghua Liu, and Jie Zhou}
\affiliation{Institute of Theoretical Physics and Department of
Physics, East China Normal University, Shanghai, 200062, China}

\date{\today}

\begin{abstract}
We study the condensation phenomenon in a zero range process on
weighted scale-free networks in order to show how the weighted
transport influences the particle condensation. Instead of the
approach of grand canonical ensemble which is generally used in a
zero range process, we introduce an alternate approach of the mean
field equations to study the dynamics of particle transport. We
find that the condensation on scale-free network is easier to
occur in the case of weighted transport than in the case of
weight-free. In the weighted transport, especially, a dynamical
condensation is even possible for the case of no interaction among
particles, which is impossible in the case of weight-free.

\end{abstract}
 \maketitle

\section{Introduction}

Condensation, a concept originally introduced by Bose and Einstein
to explain the particle condensation in momentum space, is an
intriguing phenomenon observed in real space, such as jamming in
traffic communication \cite{Zhao:2005,Liu:2006}, bunching of buses
\cite{Loan:1998}, and many mass transport models \cite{Evan:2005}.
In condensation, a finite fraction of particles may be condensed
onto a single site. Nowadays, such phenomenon is found in the
structure of complex networks, which characterize many natural and
manmade systems, such as the Internet, airline transport system,
power grid infrastructures, and the world wide web
\cite{BA:2002,rep:2006}. For example, Bianconi and Barabasi mapped
the fitness model to a Bose gas by assigning an energy to each
node and found that the fittest node can attract a finite fraction
of all links \cite{Bian:2001,Doro:2001}. Besides the static
condensation of the links of complex networks, the dynamical
condensation on scale-free (SF) networks is found recently that
the particles may completely condensed on the hub in the transport
of zero range process (ZRP) where the interaction only occurs when
the particles are stay at the same node
\cite{Noh:2005a,Noh:2005b}.

The previous studies of ZRP are mainly focused on regular lattice
\cite{Evan:2000,Evan:2005,Evan:1998}. It is found that, in the
steady state of one dimensional lattice, a finite fraction of the
total mass will condense onto a single site when the global mass
density is increased beyond a critical value. However, most of the
realistic networks are not regular but SF \cite{BA:2002,rep:2006}.
Comparing to the regular lattice, SF network is heterogeneous with
power-law degree distribution $P(k)\sim k^{-\gamma}$ and can be
characterized by the Barabasi-Albert (BA) model with $\gamma=3$
\cite{BA:1999} and its related modified models
\cite{Doro:2000,Krap:2000,Cald:2002,Goh:2001,Liu:2002}. It is
pointed out that the structure inhomogeneity is an important
factor in understanding many dynamical processes in SF networks
\cite{Vespi:2004,Sood:2005,Liu:2003}, such as rumor propagation,
virus spreading, and searching information etc. For studying the
influence of structure inhomogeneity on ZRP, Noh et al. considered
ZRP in BA model and found that the inhomogeneous structure makes
particles to condensed on the hub when the jumping rate $\delta$
is smaller than the critical point $\delta_c$
\cite{Noh:2005a,Noh:2005b}. For a specific particle, its behavior
can be considered as a random walk; and for a large number of
particles, their behaviors are a kind of diffusion processes. As
the diffusion makes the number of particles on a node fluctuate,
the usually used approach to deal with the transport statistical
properties in ZRP is the grand canonical ensemble, which can
easily give the mean particles on each node. Recently, Ref.
\cite{Evan:2005} uses the approach of canonical ensemble to
explore the condensed phase and analyze the mass distribution.
Here we introduce an alternate approach to deal with the ZRP,
i.e., the mean field equations, and find that our approach will
give the same results as that given by the approach of grand
canonical ensemble \cite{Noh:2005a,Noh:2005b} when it is used to
the same situations. Moreover, we use this approach to study the
case of weighted transport and find that the added weight may push
forward the critical value $\delta_c$ to a larger value
$\delta_c'$, and $\delta_c'$ may be even over unity!

Let's call the network with the same weight to transfer particles
on each link as {\it weight-free} network. Besides weight-free
network, there is also weighted network where each link has
different weight to transfer particles, which yields a more
realistic description of communication in real networks. Refs.
\cite{Noh:2005a,Noh:2005b} investigated how the network structure
influences particle dynamics. Our motivation here is to study how
the weight of link influences the particle dynamics. In the
weighted network, each link or node is associated with a weight
$w_i$. The weight may represent the intimacy between individuals
in social networks, or the bandwidths of routers and optical
cables in the Internet. The strength of a node, $s_i$, is defined
as the sum of all the weights of links of the node. It is found
that the strength is strongly rely on its degree with $s(k)\sim
k^{\alpha}$ \cite{Barrat:2004a,Barrat:2004b,Goh:2006} where
$\alpha$ is different constants for different networks. For
example, $\alpha$ is unity for the science cooperation network
where the weight of a link between two scientists is given roughly
by the frequency of their collaboration and $1.5$ for the
world-wide airport network where the weight is taken as the total
number of passengers of the direct flights between two connected
cities. The previous results on condensation focus on the case of
$\alpha=0$ \cite{Noh:2005a,Noh:2005b}. In this paper, we consider
the particle transport on the weighted network, i.e., the case of
$\alpha>0$. Our results show that the condensation is easier to
occur in the case of $\alpha>0$ than that of $\alpha=0$.
Especially, the condensation may occur even when the interaction
among particles does not exist, which is impossible in the case of
$\alpha=0$.

The paper is organized as follows. In Sec. II, we briefly review
the approach of grand canonical ensemble and its results of
condensation on SF networks with weight-free. Then in Sec. III, we
give the dynamical mean field equations for the weighted transport
on SF networks. Its limit will give the results of weight-free
transport. In Sec. IV, we make numerical simulations to confirm
the predictions given in Sec.III. Finally, the conclusions are
given in Sec. V.

\section{The approach of grand canonical ensemble for the case of weight-free transport}
The approach of grand canonical ensemble was first applied in the
ZRP of one-dimensional system
\cite{Evan:2005,Evan:2000,Evan:1998,Kafri:2002} and recently used
in the SF networks \cite{Noh:2005a,Noh:2005b}. Here we briefly
review the results obtained in \cite{Noh:2005a,Noh:2005b}. Suppose
$N$ particles are randomly put on a network of $L$ nodes and each
node $i$ can be occupied by any integral number of particles
$n_i=0,1,2,\cdots,N$. Because of the interaction among the
particles at the same node, some of the particles will jump out of
the node and hops to other nodes, making the particle
redistribution among $n_1, n_2,\cdots,n_L$. Hence a microscopic
configuration is represented by ${\bf n}=n_1,n_2,\cdots,n_L$. The
particles at node $i$ with $n_i>0$ may jump out with the jumping
rates $p(n_i)$ and hop from node $i$ to one of its neighbors $j$
along the link with the hopping probability $T_{j\leftarrow i}$.
In the SF networks with weight-free, $T_{j\leftarrow i}$ is taken
as $1/k_i$ if $i$ and $j$ are linked and $0$ otherwise, i.e., a
particle jumping out of the node $i$ is allowed to hop to one of
its neighboring nodes selected randomly, and the jumping rate is
taken as $p(n_i)=n_i^{\delta}$. $\delta=0$ means only one of $n_i$
particles will jump out per each time, indicating the particles
are attracting each other. In the case of $\delta=1$, all the
$n_i$ particles will jump out, implying they are moving
independently and the system reduces to a noninteracting system of
$N$ random walkers. $\delta>1$ means there is repulsive
interaction among particles of one node, while $\delta<1$ means
the attractive interaction among particles. With enough time of
evolution, the system will reach a stationary state. The mean
occupation number in the stationary state is
\begin{equation}\label{eq:occupation}
m_i(z)=x\frac{\partial ln F_{\delta}(x)}{\partial x}|_{x=zk_i}
\end{equation}
where $z$ denotes the fugacity and
\begin{equation}\label{eq:fugacity}
F_{\delta}(x)=\sum_{n=0}^{\infty}\frac{x^n}{(n!)^{\delta}}
\end{equation}

It is found that the complete condensation occurs at $\delta=0$,
where the whole fraction of particles is concentrated at the hub
\cite{Noh:2005a,Noh:2005b}. While there is no condensation at
$\delta=1$. For the case of $\delta>0$, there is a critical
\begin{equation}\label{eq:delta_c}
\delta_c=1/(\gamma-1)
\end{equation}
where the condensation will occur for $\delta\leq \delta_c$ and
does not occur when $\delta>\delta_c$. There is a crossover degree
$k_c$ for $0<\delta \leq \delta_c$, which is defined as the degree
with the average occupation number $m_i=1$. $m_i$ will be greater
than one for the nodes with $k_i>k_c$ and smaller than one for the
nodes with $k_i<k_c$. There will be a condensation when there
exists a finite $k_c>0$ and no condensation otherwise. For
$\delta\leq \delta_c$, $k_c$ can be expressed as
\begin{equation}\label{eq:cross}
k_c\sim \left\{\begin{array}{l} (ln k_{max})^{\delta_c}, \quad for \quad \delta=\delta_c \\
 (k_{max})^{1-\delta/\delta_c}, \quad for \quad \delta<\delta_c \end{array} \right.
\end{equation}
where the nodes with $k>k_c$ will be the core of condensation,
i.e., most of particles will hop to those nodes with $k>k_c$.

Furthermore, there is a scaling
\begin{equation}\label{eq:mk}
m_i=G_{\delta}(k_i/k_c),
\end{equation}
where the scaling functions behaves as $G_{\delta}(y<<1)\sim y$
and $G_{\delta}(y)(y\geq 1)\sim y^{1/\delta}$.

\section{Dynamical mean field equations for the case of weighted transport}
According to the strength distribution $s(k)\sim k^{\alpha}$
\cite{Barrat:2004a,Barrat:2004b}, the strength is a nonlinear
function of $k$ and the nodes with the same $k$ have the same
strength. The strength will be different for the $k_i$ neighbors
of node $i$. When a jumping out particle hops to one of its
neighbors, it will not choose all the links of the node on equal
footing but with different rates to different neighbors.
Therefore, we here assume the hopping rate $T_{j\leftarrow i}\sim
k_j^{\alpha}$. Making renormalization, we have
\begin{equation}\label{eq:jump1}
T_{j\leftarrow i}=\frac{k_j^{\alpha}}{\sum_{j'\in
B_i}k_{j'}^{\alpha}},
\end{equation}
where $B_i$ denotes the set of neighbors of node $i$ and Eq.
(\ref{eq:jump1}) will return to the case of weight-free when
$\alpha=0$.

We still suppose here that the jumping rates $p(n_i)$ satisfy
$p(n_i)=n_i^{\delta}$ and restrict $\delta$ to $0\leq \delta \leq
1$. With the evolution, $n_i$ will change with time and may be
different for different nodes with the same $k$. For using the
mean field approach, we transfer the description of $n_i$ for each
node to the description of the mean occupation number $m_k(t)$ for
the nodes with the same $k$, i.e., $m_k(t)$ is the average of all
the $n_i$ of the nodes with degree $k$. Hence $m_k(t)$ is no
longer necessary to be an integer. Correspondingly, we transfer
the jumping rate $p(n_i)$ to $p(m_k)=m_k^{\delta}$. After this
transformation, let's see how many particles of a node can jump
out at each time step. When $m_k\leq 1$, $m_k^{\delta}\geq m_k$
and the jumping rate is in fact $m_k$ but not $m_k^{\delta}$ as we
don't have so many average particles $m_k^{\delta}$ at those nodes
with degree $k$. When $m_k>1$, $m_k^{\delta}<m_k$ and hence the
jumping rate is $m_k^{\delta}$. Except the aspect of jumping out,
at the same time, a node accepts particles from its neighbors. The
incoming particles can be classified into two parts: One from the
node with $m_k<1$ and the other from the nodes with $m_k\geq 1$.
The incoming particles from one neighboring node with degree $k'$
is $P(k'|k)m_{k'}(t)k^{\alpha}/\sum_{j\in B_{k'}}k_j^{\alpha}$
when $m_{k'}(t)<1$ and
$P(k'|k)m_{k'}^{\delta}(t)k^{\alpha}/\sum_{j\in
B_{k'}}k_j^{\alpha}$ when $m_{k'}(t)\geq 1$, where $P(k'|k)$ is
the conditional probability for a link which represents the
possibility for a node with degree $k$ to connect a node with
degree $k'$. We use $k_0$ and $k_{max}$ to represent the minimum
and maximum degree of the network, respectively, and use $k_c(t)$
to denote the degree for $m_k(t)=1$. In the BA model, $k_0$ is a
constant and $k_{max}\sim L^{\beta}$ with $\beta=1/(\gamma-1)$. By
all these quantities, we have the mean field equations for the
evolution of $m_k(t)$
\begin{eqnarray}\label{eq:mean-field}
\frac{\partial m_k(t)}{\partial
t}&=&-m_k(t)+k(\sum_{k_0}^{k_c}p(k'|k)m_{k'}(t)
\frac{k^{\alpha}}{\sum_{j\in
B_{k'}} k_j^{\alpha}}+\nonumber \\
& &\sum_{k_c}^{k_{max}}p(k'|k)m_{k'}^{\delta}(t)
\frac{k^{\alpha}}{\sum_{j\in
B_{k'}} k_j^{\alpha}}), m_k(t)<1 \nonumber \\
\frac{\partial m_k(t)}{\partial
t}&=&-m_k^{\delta}(t)+k(\sum_{k_0}^{k_c}p(k'|k)m_{k'}(t)
\frac{k^{\alpha}}{\sum_{j\in
B_{k'}} k_j^{\alpha}}+\nonumber \\
& &\sum_{k_c}^{k_{max}}p(k'|k)m_{k'}^{\delta}(t)
\frac{k^{\alpha}}{\sum_{j\in B_{k'}} k_j^{\alpha}}), m_k(t)\geq 1
\end{eqnarray}
where $\sum_{j\in B_{k'}} k_j^{\alpha}$ denotes the sum to all the
neighbors of the node with degree $k'$. Obviously, $m_k(t)$
depends on $p(k'|k)$. For getting the detailed form of $p(k'|k)$,
here we consider the BA model as the underlying network. As the BA
model is non-assortative mixing, its conditional probability
satisfies $p(k'|k)=k'p(k')/<k>$ \cite{Newman:2002,Catan:2005}.
Borrowing the definition of the average degree of the nearest
neighbors of a node with degree $k$, i.e.,
$\bar{k}_{nn}(k)=\sum_{k'}k'p(k'|k)$, we have the average
\begin{eqnarray}\label{eq:nnalpha}
\bar{k}_{nn}^{\alpha}(k)&=&\int_{k_0}^{k_{max}}p(k'|k)k'^{\alpha}dk' \nonumber \\
&=& <k^{\alpha+1}>/<k>
\end{eqnarray}
It is easy to see that $\bar{k}_{nn}^{\alpha}$ does not depend on
the degree $k$. Hence, we have $\sum_{j\in B_{k'}}
k_j^{\alpha}=k'\bar{k}_{nn}^{\alpha}$ and Eq.
(\ref{eq:mean-field}) becomes
\begin{eqnarray}\label{eq:mean-field1}
\frac{\partial m_k(t)}{\partial
t}&=&-m_k(t)+k^{\alpha+1}A(t), \quad m_k(t)<1 \nonumber \\
\frac{\partial m_k(t)}{\partial
t}&=&-m_k^{\delta}(t)+k^{\alpha+1}A(t), \quad m_k(t)\geq 1
\end{eqnarray}
where $A(t)=(\sum_{k_0}^{k_c}p(k')m_{k'}(t)
+\sum_{k_c}^{k_{max}}p(k')m_{k'}^{\delta}(t))/(<k>\bar{k}_{nn}^{\alpha})$.

Eq. (\ref{eq:mean-field1}) is the evolution equation of particle
and will reach the stationary state when it evolves enough time.
In the stationary state, we may solve Eq. (\ref{eq:mean-field1})
by $\partial m_k(t)/\partial t=0$ to get the stabilized $m_k$. By
doing this we have
\begin{eqnarray}\label{eq:stationary}
m_k &=& k^{\alpha+1}A, \quad m_k<1 \nonumber \\
m_k &=& (k^{\alpha+1}A)^{1/\delta}, \quad m_k\geq 1.
\end{eqnarray}
where $A=(\sum_{k_0}^{k_c}p(k')m_{k'}
+\sum_{k_c}^{k_{max}}p(k')m_{k'}^{\delta})/(<k>\bar{k}_{nn}^{\alpha})$
does not depend on the time $t$. We have $A=k_c^{-(\alpha+1)}$
from the condition $m_k=1$ for $k=k_c$. Instituting $A$ into Eq.
(\ref{eq:stationary}) we have
\begin{eqnarray}\label{eq:stationary2}
m_k &=&(k/k_c)^{\alpha+1}, \quad m_k<1 \nonumber \\
m_k &=& (k/k_c)^{(\alpha+1)/\delta}, \quad m_k\geq 1.
\end{eqnarray}
Obviously, Eq. (\ref{eq:stationary2}) will go back to Eq.
(\ref{eq:mk}) when $\alpha=0$.

Recall that in the case of one dimensional lattice, one necessary
condition for the condensation is that its density $\rho=N/L$
should be larger than a critical value $\rho_c>0$. While in the BA
model of weight-free, there is no $\rho_c$ for $\rho$. The
condensation occurs at any finite value of the particle density,
i.e., $\rho_c=0$, and the particles can completely condensed at
the hub or a few high-degree nodes \cite{Noh:2005a,Noh:2005b}.
Here the limit case of $\alpha=0$ is the BA model of weight-free,
hence we expect $\rho_c=0$ for $\alpha>0$, which will be
demonstrated later. For confirming it, in the case of weighted
transport we set the particle density $\rho<1$. In the limit of
$L, N\rightarrow \infty$ we have $m_k<1$ for all nodes if there is
no condensation; otherwise, part of $m_k$ will become greater than
unity. Therefore, the $k_c$ for $m_k=1$ is nothing but the
crossover degree. The number of condensed particles is
\begin{equation}\label{eq:density}
N_{con}\sim N\int_{k_c}^{k_{max}}
(k/k_c)^{(\alpha+1)/\delta}p(k)dk
\end{equation}

In the condensed state, the nodes with $k_c\leq k\leq k_{max}$
have the capacity to take an infinite number of particles. The
non-divergence of the integration in Eq. (\ref{eq:density}) gives
\begin{equation}\label{eq:critical}
\delta_c'=\frac{\alpha+1}{\gamma-1}
\end{equation}
The condensation will occur for the case of $\delta<\delta_c'$.
From Eq. (\ref{eq:critical}) one can see that in the case of
$\alpha=0$, $\delta_c'=1/(\gamma-1)$ returns to the case of
weight-free of Eq. (\ref{eq:delta_c}). By $N_{con}/N\sim O(1)$, we
have
\begin{equation}\label{eq:cross1}
k_c^{(\alpha+1)/\delta}\sim \int_{k_c}^{k_{max}}
k^{(\alpha+1)/\delta}k^{-\gamma}dk,
\end{equation}
and further we have
\begin{equation}\label{eq:cross2}
k_c\sim \left\{\begin{array}{l} (ln k_{max})^{\delta_c}, \quad for \quad \delta=\delta_c' \\
 (k_{max})^{1-\delta/\delta_c'}, \quad for \quad \delta<\delta_c' \end{array} \right.
\end{equation}
Obviously, it has the same form with Eq. (\ref{eq:cross}) except
$\delta_c$ is replaced by $\delta'$. From Eq. (\ref{eq:cross2})
one can see that $k_c$ does not depend on the particle density
$\rho$, indicating the condensation can occur at any $\rho$.
Therefore, we also have $\rho_c=0$ for the weighted transport.

For the situation of $\delta<\delta_c'$, from Eq.
(\ref{eq:stationary2}) the particles accumulated at the hub is
\begin{eqnarray}\label{eq:hub1}
m_{hub}&\sim& (k_{max}/k_c)^{(\alpha+1)/\delta} \nonumber\\
&\sim& (k_{max}^{\delta/\delta_c'})^{(\alpha+1)/\delta} \sim L,
\end{eqnarray}
indicating $m_{hub}$ scales linearly with $L$ and hence there is
condensation at the hub. And for the situation of
$\delta>\delta_c'$, the particles at the hub is
\begin{equation}\label{eq:hub2}
m_{hub}\sim k_{max}^{(\alpha+1)/\delta}\sim L^{\delta_c'/\delta},
\end{equation}
indicating $m_{hub}$ scales sublinearly with $L$ and hence there
is no condensation. When there is condensation, from Eq.
(\ref{eq:stationary2}) we have
\begin{equation}\label{eq:condense}
m_k\sim k^{\eta}
\end{equation}
with $\eta=(\alpha+1)/\delta$ for $\delta<\delta_c'$ and $m_k>1$.
The scaling $\eta$ will decrease to
$(\alpha+1)/\delta_c'=\gamma-1$ when $\delta$ increase to
$\delta_c'$, which gives a critical $\eta_c=2$ for $\gamma=3$.
Therefore, we can judge the occurrence of condensation by checking
the scaling $\eta$ for $m_k>1$, i.e., there is condensation if
$\eta>\gamma-1$ and no condensation otherwise.

Moreover, from Eq. (\ref{eq:critical}) we can see that, for BA
model with $\gamma=3$, it is possible to have $\delta_c'>1$ for
$\alpha>1$. In this case the condensation will occur for the whole
range of $0\leq \delta\leq 1$, which is impossible in the case of
weight-free. For example, taking $\delta=1$ for the case of
weight-free, all the particles at node $i$ will jump out of the
node per each time, resulting a uniform distribution of particles
and hence no condensation. However, in the case of weighted
transport, the situation is totally changed. During the evolution,
all the particles will jump out at each time and hop with the
favorite to one of the neighbors with the largest degree.
Gradually, most of the particles will move to the larger and
larger nodes, and finally to the hub around. After that, all the
particles at the hub will jump out per each time when $\delta=1$,
but at the same time, most of the particles of the hub's
neighboring nodes will hop to the hub, resulting in a dynamical
equilibrium. Namely, $\delta=1$ for $\delta_c'>1$ is a kind of
completely {\sl exchange} condensation, which is different from
the case of $\delta<1$ where most of the previous particles at the
hub will still stay at the hub. This kind of complete exchange
condensation also exists for the case of $1<\delta<\delta_c'$
where the jumping rate $p(m_k)=m_k$ is the same with that of
$\delta=1$.

\section{Numerical simulations}
\begin{figure}
\epsfig{figure=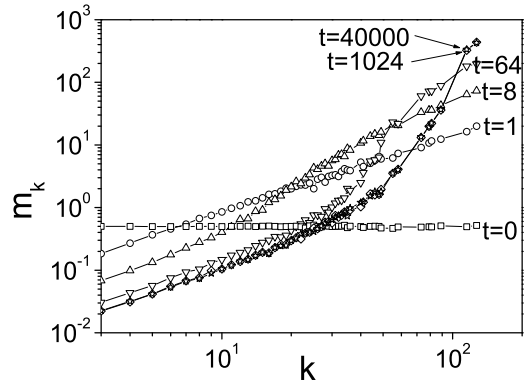,width=0.9\linewidth}\vspace{-0.5cm}
\caption{Evolution of particle distribution in weighted transport
for an arbitrary $\alpha=0.4$ and $\delta=0.2<\delta_c=0.7$. }
\label{evolution}
\end{figure}

In numerical simulations, we first construct a BA growing network
with the size $L=2000$, the average degree $<k>=6$, and the degree
distribution $P(k)\sim k^{-3}$ according to the algorithm given in
Ref. \cite{BA:1999,Liu:2002}. Then we put $N=1000$ particles at
the $L$ nodes randomly, i.e., the particle density $\rho=0.5$.
Hence $m_k(t=0)$ is uniform for different $k$, see the horizontal
line for the case of $t=0$ in Fig. \ref{evolution}. At each time
step, we let the $N$ particles evolve according to the jumping
rate $p(n_i)$ and the hopping rate $T_{j\leftarrow i}$ in Eq.
(\ref{eq:jump1}). Namely, we choose $min(n_i, n_i^{\delta})$
particles from the $n_i$ particles of node $i$ and let them hop to
the neighbors of node $i$ according to Eq. (\ref{eq:jump1}), where
$min(n_i, n_i^{\delta})$ means taking the smaller one from $n_i$
and $n_i^{\delta}$. By this way, we find that the numerical
simulations completely confirm the evolution Eq.
(\ref{eq:mean-field1}) and the stationary distribution Eq.
(\ref{eq:stationary2}). For example, from Eq.
(\ref{eq:mean-field1}) we have $m_k(t=1)=A(0)k^{\alpha+1}\sim
k^{\alpha+1}$, indicating a power-law relation between $m_k$ and
$k$ for all the $k$ at $t=1$. Our numerical simulation has
confirmed it, see the straight line for the case of $t=1$ in Fig.
\ref{evolution} where the parameters are taken as $\alpha=0.4$ and
$\delta=0.2<\delta_c'=0.7$. This algebraic relation will be kept
until the emergence of some $m_k(t)>1$. After that, we know from
Eq. (\ref{eq:mean-field1}) that the nodes with $m_k(t)<1$ for
small $k$ will keep the algebraic relation $m_k(t)\sim
A(t-1)k^{\alpha+1}$ but the nodes with $m_k(t)>1$ for large $k$
will depends on both $k$ and $m_k(t-1)$ because of the
nonlinearity of the first term $m_k^{\delta}(t)$ in Eq.
(\ref{eq:mean-field1}). The numerical confirmation is given by the
curves for the cases of $t=8$ and $64$ in Fig. \ref{evolution}. It
is easy to see from these two curves that the part with smaller
$k$ has the same slope with that of the curve of $t=1$ and the
part with larger $k$ is complicated. After further evolution, the
system will reach its stationary state. The lowest lines in Fig.
\ref{evolution} show the numerical results for $t=1024$ and
$40000$ respectively. Obviously, the case of $t=1024$ is
overlapped with the case of $t=40000$, indicating the stationary
state is reached.

\begin{figure}
\epsfig{figure=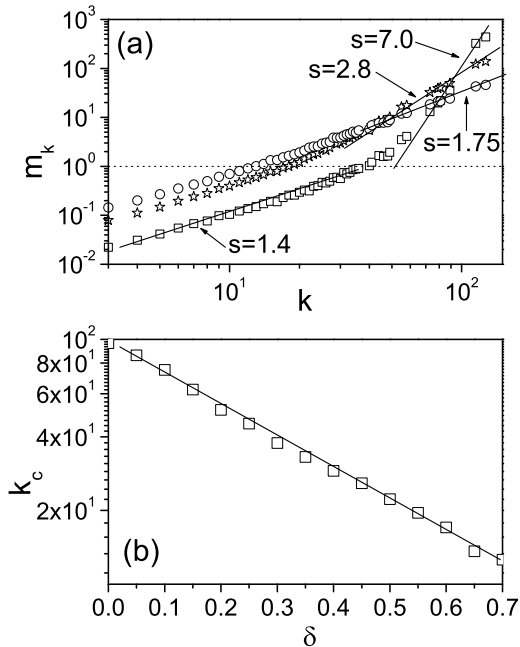,width=0.9\linewidth}\vspace{-0.5cm}
\caption{(a) Stabilized $m_k$ versus $k$ for $\alpha=0.4$ and
$\delta_c'=0.7$ where the lines with ``squares", ``stars", and
``circles" denote $\delta=0.2$, $0.5$, and $0.8$, respectively.
The drawn lines for slope $s=1.4, 7.0, 2.8$, and $1.75$ are for
reference. (b) Crossover degree $k_c$ versus $\alpha$.}
\label{cross}
\end{figure}

In the stationary state, from Eq. (\ref{eq:stationary2}) we know
that the stabilized $m_k$ will have two slopes and both depends on
the weight scaling $\alpha$, i.e., slope $\alpha+1$ for $m_k<1$
and slope $(\alpha+1)/\delta$ for $m_k>1$. The numerical
simulations have completely confirmed this. Fig. \ref{cross}(a)
shows the results for $\alpha=0.4$ and $\delta=0.2$, $0.5$, and
$0.8$ respectively. By Eq. (\ref{eq:critical}) we get
$\delta_c'=0.7$, so the case of $\delta=0.2$ and $0.5$ should have
condensation and the case of $\delta=0.8>\delta_c'$ no
condensation. This can be confirmed by checking the scaling $\eta$
in Eq. (\ref{eq:condense}). It is easy to see from Fig.
\ref{cross}(a) that the slopes are $7.0$ and $2.8$ for the cases
of $\delta=0.2$ and $0.5$, which are greater than $\eta_c=2$, and
$1.75$ for the case of $\delta=0.8$, which is less than $2$. These
values also confirm the prediction of Eq. (\ref{eq:stationary2})
in which we have the slopes $\alpha+1=1.4$, and
$(\alpha+1)/\delta=7, 2.8$, and $1.75$ for $\delta=0.2, 0.5$, and
$\delta=0.8$, respectively, see the solid reference lines with
different slopes $s$ in Fig. \ref{cross}(a). On the other hand,
from Eq. (\ref{eq:cross2}) we have $ln k_c\sim C\delta$. The
crossover degree $k_c$ can be measured from the intersection
between the dotted line $m_k=1$ and the measured curves in Fig.
\ref{cross}(a). By this way we can get $k_c$ for different
$\delta$. Fig. \ref{cross}(b) shows the result. It is easy to see
that Fig. \ref{cross}(b) is a straight line with negative slope,
which qualitatively explain Eq. (\ref{eq:cross2}).

\begin{figure}
\epsfig{figure=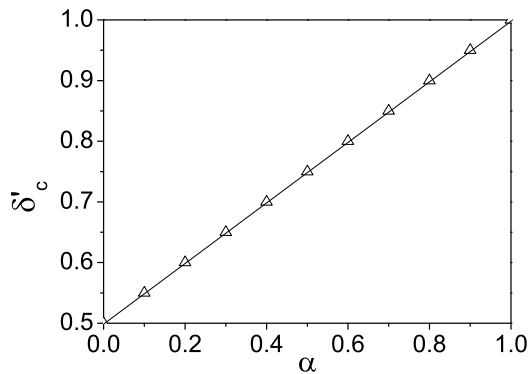,width=0.9\linewidth}\vspace{-0.5cm}
\caption{The critical parameter $\delta_c'$ of condensation versus
the weight parameter $\alpha$. } \label{critical}
\end{figure}

By the fact that $\eta>\eta_c$ means condensation, we may
determine $\delta_c'$ numerically. For a given $\alpha$, we
increase $\delta$ gradually and check $\eta$ for each $\delta$.
The critical $\delta_c'$ can be identified as the value that
separates the $\eta>2$ and $\eta<2$. By this way, we obtain
$\delta_c'$ for different $\alpha$. Fig. \ref{critical} shows the
result. Obviously, it is a straight line with slope $0.5$ and thus
confirms Eq. (\ref{eq:critical}). When $\alpha$ is over unity,
however, we cannot obtain $\delta_c'$ numerically as $\delta$ is
limited to unity for the jumping rate $p(n)$.

\begin{figure}
\epsfig{figure=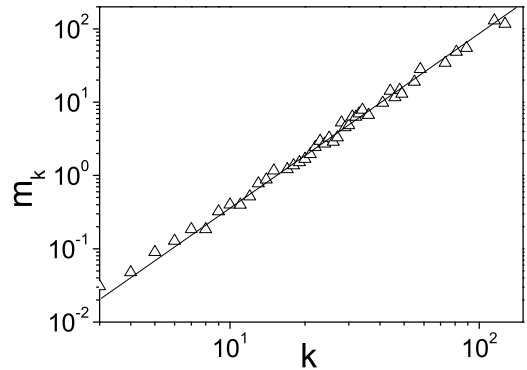,width=0.9\linewidth}\vspace{-0.5cm}
\caption{The stabilized $m_k$ versus $k$ for $\alpha=1.4$ and
$\delta=1<\delta_c'=1.2$. } \label{repulsive}
\end{figure}

Furthermore, from Eq. (\ref{eq:stationary2}) we can see that the
two slopes for $m_k<1$ and $m_k>1$ will become the same when
$\delta=1$. This phenomenon is also confirmed by our numerical
simulation. Fig. \ref{repulsive} shows the result where the
measured slope $2.37$ (the solid line) is consistent with the
theoretical prediction $\alpha+1=(\alpha+1)/\delta=2.4$ and the
fact that $2.37>2$ confirms the condensation for
$\delta=1<\delta_c'$.

\section{Discussions and conclusions}
The condensation in ZRP was usually discussed in mass transport
and only a little attention had been paid to the interacting
dynamical systems on SF networks. The work of Noh et al. makes ZRP
on SF networks become reality. As complex network has very rich
features, Noh's work is very important but only the beginning of
ZRP on complex network. There are a plenty of factors in networks
which may affect the condensation, such as the distribution,
clustering coefficient, assortativity, and weight etc. This paper
investigates one of the factors. Our result show that the weight
may significantly influence the condensation and make it possible
for the condensation to occur even for $\delta=1$ when
$\delta_c'>1$, which is impossible in the weight-free network. We
are continue along this way.

 In conclusions, we have discussed the dynamical
condensation on SF network with weighted transport. We have
introduced a set of dynamical mean-field equations in ZRP to
describe the evolution of particle and find that the particle
condensation depends on not only the jumping rate, but also the
weight of transport. Our results show that the mean field approach
is completely equivalent to the grand canonical ensemble in the
situation of $\alpha=0$ and the condensation is easier to occur in
the weighted network than that in the network with weight-free.
The critical value of condensation in weighted network is
$\delta_c'=(\alpha+1)/(\gamma-1)$, which is larger than the value
$\delta_c=1/(\gamma-1)$ of weight-free.

This work was supported by the NNSF of China under Grant No.
10475027, by the PPS under Grant No. 05PJ14036, by SPS under Grant
No. 05SG27, and by NCET-05-0424.

\end{document}